\documentclass
[preprint,prb,a4paper,showpacs,amsfonts,amssymb,final,notitlepage,11pt,tightenlines]{revtex4}%
\usepackage{amsmath}
\usepackage{graphicx}
\usepackage{amsfonts}
\usepackage{amssymb}%
\begin{document}
\author{M. El Massalami, R. E. Rapp, F. A. B. Chaves}
\affiliation{Instituto de Fisica-UFRJ, CxP 68528, 21945-970, Rio de Janeiro,Brazil}
\author{H. Takeya,}
\affiliation{National Institute for Materials Science,1-2-1 Sengen, Tsukuba,Ibaraki,
305-0047, Japan,}
\author{C. M. Chaves,}
\affiliation{Departamento de Fisica, PUC-Rio, 22453-900 Rio de Janeiro, Brazil}
\title{Magnon Specific Heat of Single Crystal Borocarbides $R$Ni$_{2}$B$_{2}$C
($R$=Tm, Er, Ho, Dy, Tb, Gd).}
\date{\today{}}

\begin{abstract}
Zero-field specific heat of the single crystals $R$Ni$_{2}$B$_{2}$C ($R$= Er,
Ho, Dy, Tb, Gd) was measured within the temperature range 0.1 K$<T<$25 K.
\ Linearized spin wave analysis was successfully applied to account for and to
rationalize the thermal evolution of the low--temperature magnetic specific
heats of all the studied compounds (as well as the one reported for TmNi$_{2}%
$B$_{2}$C) in terms of only two parameters, namely an energy gap $\Delta$ and
a characteristic temperature $\theta$. The evolution of $\theta$ and $\Delta$
across the studied compounds correlates very well with the known magnetic
properties. $\theta,$ as a measure of the effective $RKKY$ exchange
couplings,\ scales reasonably well with the de Gennes factor. $\Delta,$ on the
other hand, reflects predominately the anisotropic properties: $\sim2$ K for
GdNi$_{2}$B$_{2}$C, $\sim6$ K for ErNi$_{2}$B$_{2}$C, $\sim7$ K for TbNi$_{2}%
$B$_{2}$C, and $\sim8$ K\ for each of HoNi$_{2}$B$_{2}$C and DyNi$_{2}$B$_{2}%
$C. The equality in $\Delta$ of HoNi$_{2}$B$_{2}$C and DyNi$_{2}$B$_{2}$C,
coupled with the similarity in their magnetic configurations, indicates that a
variation of $x$ in the solid solution Ho$_{x}$Dy$_{1-x}$Ni$_{2}$B$_{2}$C
($x<0.8$ and $T_{c}<T_{N}$) would not lead to any softening of $\Delta$. This
supports the hypothesis of Cho et al (PRL \textbf{77},163(1996)) concerning
the influence of the collective magnetic excitations on the superconducting
state. This work underlines the importance of \ spin-wave excitations for a
valid description of low-temperature thermodynamics of borocarbides.

\end{abstract}
\maketitle

\section{Introduction}

Wide varieties \ of magnetic ground structures are manifested in $R$Ni$_{2}%
$B$_{2}$C\ series (see e.g.
Refs.\cite{Lynn-RNi2B2C-mag-xrd,Detlefs-GdNi2B2C-XRES} and Tab.I): members
with $R=$Ho, Dy, Nd, Pr adopt commensurate antiferromagnetic (AF) arrangements
while those with $R=$Tm, Er, Tb, Gd, assume squared-up spin-density wave (SDW)
states. These ground structures are stabilized by a\ fine balance of exchange,
crystalline electric field (CEF), and magnetoelastic forces. Under magnetic
field or temperature variation, most of these structures undergo a cascade of
phase transformations, yielding rich varieties of field-temperature ($H-T$)
phase diagrams ( see e.g.
Refs.\cite{Canfield-Ho-HT-Diagram,Budko-Er-HT-diagram-2,Detlefs-Ho-HT-Diagram,Canfield-Dy-HT-diagram}%
). \ Interestingly, most of the zero-field magnetic states of $R$=Tm, Er, Ho,
Dy coexist with superconductivity, presenting model compounds wherein the
interplay between superconductivity and magnetism can be investigated. Such
investigations revealed that the superconductivity, though much influenced by,
has a very weak influence on the prevailing magnetic order:\cite{Amici-thesis}
the energy gain due to the onset of magnetic order dominates by two order of
magnitudes over that due to the onset of superconductivity.

The zero-field part of the $H-T$ phase diagram of HoNi$_{2}$B$_{2}$C is
particularly interesting: superconductivity sets-in at $T_{c}\simeq8$ K. Just
below this point, an incommensurate spiral state, $\overrightarrow{k_{c}%
}\simeq$.92$\overset{\ast}{c}$, develops (see e.g.
Refs.\cite{Lynn-RNi2B2C-mag-xrd,Canfield-Ho-HT-Diagram,Canfield-HoNi2B2C-M-Cp}%
). Furthermore, at $\sim$6.3 K, an additional modulated state with
$\overrightarrow{k_{a}}\simeq$0.55$\overset{\ast}{a}$ emerges and around 5 K a
deep minimum in $H_{c2}$ develops.\cite{Canfield-RNi2B2C-Hc2} At $T_{N}=$5 K,
an orthorhombic lattice distortion \cite{Kreyssig-Ho-distortion} sets-in and,
concomitantly, both the spiral and the a-axis modulated states are replaced by
a commensurate AF structure which coexists with superconductivity down to the
lowest measurable temperatures. Generalized susceptibility calculation
\cite{Joo-generalized-susc} related these modulated state to maxima in the
exchange coupling transform,$\ J(k)$.

Remarkably, the magnetic ground structures of the heavy and magnetic
$R$Ni$_{2}$B$_{2}$C compounds (see Tab.I) are particularly simple: in spite of
the manifestation of an orthorhombic distortion and a
liquid-helium-temperature magnetic modulation, their ground structures are
either an equal-amplitude, AF-type squared-up state (as in $R=$Tm, Er, Tb, Gd)
or an equal-amplitude, colinear, commensurate AF state (as in $R=$Ho, Dy).
Then,\ it is of interest to investigate whether the low-temperature
thermodynamics of these $R$Ni$_{2}$B$_{2}$C can be described in terms of
\ small-amplitude spin-wave excitations and, in addition, to elucidate the
character and dimensionality of \ these excitations. These magnetic
excitations can be probed by various techniques, among which is the magnetic
specific heat. We carried out extensive zero-field specific heat measurements
on five single crystals $R$Ni$_{2}$B$_{2}$C ($R$= Er, Ho, Dy, Tb, Gd) covering
at least the range $0.1$ K$<T<T_{N}.$ These specific heats, together with
\ that of TmNi$_{2}$B$_{2}$C (Ref.\cite{Movshovich-TmNi2B2C}), reveal a
diversified and wide varieties of thermal evolutions. None the less, based on
a phenomenological model, all of the specific heat curves can be systematized
in terms of only two parameters, namely an effective exchange coupling and a
magnetic anisotropy interaction.

The format of this paper is as follows: In Sec.II, we derive an approximate,
but of wide applicability, expression for the magnon specific heat.
Experimental techniques and procedures are described in Sec.III. Results and
their analysis are described in Sec.IV and discussed in Sec.V.

\section{Approximate expressions for magnon specific heat of $R$Ni$_{2}$%
B$_{2}$C.}

The\ magnetic structures of $R$Ni$_{2}$B$_{2}$C ($R$=Tm-Gd) can be visualized
as magnetic layers that are stacked along the
c-axis.\cite{Lynn-RNi2B2C-mag-xrd} The most \ dominant interactions are the
RKKY and anisotropic couplings. The former can be approximated by effective
isotropic couplings while the latter (a combination of \ dominant CEF and
weaker dipolar and anisotropic exchange forces) by an easy-axis anisotropy
field $\vec{H}_{a}$:\cite{Kittel-SpinWave} this $\vec{H}_{a}$ representation
is convenient for spin-wave calculation and is valid for the low-temperature
orthorhombic-distorted equal-amplitude collinear AF or squared-up phases of
borocarbides (see Sec.V). Considering the above mentioned magnetic arrangement
and the crystal symmetry, the magnetic couplings can be conveniently divided
into two classes: $J_{ij}^{A}$ that couples moments $i$ and $j$ within the
same layer (denoted as $A$ or $B$) and $J_{ij}^{AB}$ that couples moments from
different layers. Then at zero external field, the following Hamiltonian is
expected to capture most of their low-temperature \ properties:%

\begin{equation}
\mathcal{H}=-\sum_{<ij>,L\in A,B}J_{ij}^{L}\overrightarrow{S_{i}^{L}%
}.\overrightarrow{S_{j}^{L}}+\sum_{<ij>A,B}J_{ij}^{AB}\overrightarrow
{S_{i}^{A}}.\overrightarrow{S_{j}^{B}}-g\mu_{B}\vec{H_{a}}\sum_{j\in
A}\overrightarrow{S_{j}^{A}}+g\mu_{B}\vec{H_{a}}.\sum_{j\in B}\overrightarrow
{S_{j}^{B}} \label{H}%
\end{equation}
The first and second sum represent, respectively, interactions within the same
layer and among different layers. The last two terms represent the anisotropic
interactions. By standard spin-wave analysis, we obtained the following
dispersion relation:
\begin{equation}
\hbar\omega_{k}=\sqrt{\left[  SJ_{\bot}(0)-SJ_{\bot}(k)+SJ_{\Vert}(0)+g\mu
_{B}H_{a}\right]  ^{2}-\left[  SJ_{\Vert}(k)\right]  ^{2}} \label{dispersion}%
\end{equation}
where $J(k)=\sum J_{ij}\exp(ik(r_{i}-r_{j})).$ $J_{\Vert}$ ($J_{\bot}$)
represents the Fourier-transform parallel (perpendicular) to the c-axis. The
energy gap, $\Delta=\hbar\omega_{k=0},$ is:
\begin{equation}
\Delta=\sqrt{\left(  g\mu_{B}H_{a}\right)  ^{2}+2J_{\Vert}(0)g\mu_{B}SH_{a}}
\label{gap}%
\end{equation}
which defines an AF resonance frequency similar to the uniform mode of
ordinary
AFs.\cite{Kubo-SpinWave,Eisele-Keffer-SpinWave,Kittel-SpinWave,Walker-SpinWave}
Evidently, (i) $\Delta$ is zero whenever there is no anisotropy and (ii)
$\Delta$ does not depend on the type nor the strength of the intralayer coupling.

For evaluating the magnon specific heat (or other thermodynamical quantities),
an explicit expression of $J(k)$ is required. In the absence of \ such an
expression and for low-temerpature range, it is a common practice to assume a
long-wave limit. Here, we restricted the expansion of \ $J(k)$ to the nearest
neighbors only, leading to:
\begin{equation}
\hbar\omega_{k}\approx\sqrt{\Delta^{2}+C_{x}k_{x}^{2}+C_{y}k_{y}^{2}%
+C_{z}k_{z}^{2}}\text{ } \label{dispersion-quadratic}%
\end{equation}
where $C_{x}$ ($\approx C_{y}$ for weak orthorhombic distortion) and $C_{z}$
are functions of the exchange couplings and geometrical factors ($a$ and $c$
are unit-cell parameters):%
\begin{align}
C_{x}  &  =16(J_{\bot}+J_{\Vert})J_{\Vert}S^{2}a^{2}+2J_{\bot}S(g\mu_{B}%
H_{a})a^{2}\label{C-definition}\\
C_{z}  &  =16\ J_{\Vert}^{2}S^{2}c^{2}\nonumber
\end{align}
Then, the zero-field magnon specific heat is (rewritten so as to conform with
the notation of Ref.\cite{Eisele-Keffer-SpinWave}):%

\begin{equation}
C_{M}(T)=3^{\frac{3}{2}}R\left(  \Delta/\theta\right)  ^{3}\left(
\Delta/2T\pi^{2}\right)  \sum_{m=1}^{\infty}\left[  BesselK(2,m\Delta
/T)+BesselK(4,m\Delta/T)\right]  \label{Cm-bessel}%
\end{equation}
where $BesselK$ represents the modified Bessel function and
\begin{equation}
\theta=z\left\vert J_{eff}\right\vert S=\sqrt[3]{3^{\frac{3}{2}}2(C_{x}%
C_{y}C_{z})^{\frac{1}{2}}/(a^{2}c)} \label{Theta}%
\end{equation}
is a characteristic temperature, based on which $\left\vert J_{eff}\right\vert
$ can be defined as being an effective exchange interaction that couples the
magnetic moment to its $z$ nearest neighbors. Notice that both
Eq.\ref{Cm-bessel} and Eq.\ref{Theta} account as well for the weakly
orthrohmic-distorted state.

For $T<\Delta$, Eq.\ref{Cm-bessel} reduces to the exponential form:
\begin{equation}
C_{M}(T)\simeq2\frac{3^{\frac{3}{2}}R\Delta^{\frac{7}{2}}}{\pi^{\frac{3}{2}%
}\theta^{3}T^{\frac{1}{2}}}\exp(-\Delta/T) \label{Cm-AF-lowT}%
\end{equation}
while for isotropic compounds or $T\gg\Delta$, it reduces to the
high-temperature limit: \cite{Kubo-SpinWave,Eisele-Keffer-SpinWave}
\begin{equation}
C_{M}(T)\simeq\frac{3^{\frac{3}{2}}4\pi^{2}}{15}R\left(  T/\theta\right)  ^{3}
\label{Cm-AF-highT}%
\end{equation}
in this case, Eq.\ref{Theta}\ highlight the useful definition of $\left|
J_{eff}\right|  $.

A long-wave dispersion relation for an isotropic quasi-\textit{2}$d$ case can
be derived from Eq.\ref{dispersion-quadratic}, if we set $C_{z}\ll C_{x}$
($\left\vert J_{\Vert}\right\vert \ll J_{\bot}$) and $H_{a}=0$:
\begin{equation}
\hbar\omega_{k}\approx8J_{\Vert}S+J_{\bot}Sa^{2}(k_{x}^{2}+k_{y}^{2})
\label{dispersion-2d}%
\end{equation}
Then, to lower order in $8J_{\Vert}S/T$, one obtains $C_{M}(T)=\pi
RT/12SJ_{\bot}$ which reproduces the leading linear-in-$T$ term in the
expression reported by Movshovich et al \ \cite{Movshovich-TmNi2B2C} who
(starting from a quadratic dispersion relation and including correction for
the 2d and the magnon-magnon interaction) obtained for the range$\ \left\vert
J_{\Vert}\right\vert S<T<T_{N}$:%
\begin{equation}
C_{M}(T)=(\pi R/12)(T/SJ_{\bot}-6J_{\Vert}/\pi^{2}J_{\bot}+4J_{\Vert}%
S/3\pi^{2}J_{\bot}T) \label{Cm-2d}%
\end{equation}

It is worth remarking that Eqs.\ref{H}-\ref{dispersion-quadratic}%
,\ref{Cm-bessel} are of a more wide applicability than our above analysis
might have suggested. Furthermore, depending on only $\theta$ and $\Delta$, a
variety of expressions for the magnetic specific heat can be derived; each
admitting different limits for its thermal evolution depending on the
relations among $T$ and $\Delta$ (compare Eq.\ref{Cm-bessel} with the limit
Eqs.\ref{Cm-AF-lowT},\ref{Cm-AF-highT},\ref{Cm-2d}). Based on such a scheme,
one is capable of rationalizing the vast varieties of \ the\ low-temperature
thermal evolution of thermodynamical quantities (such as $C_{M}(T)$)
encountered in these (and any series similar to) borocarbides. It is reminded
that this analysis is not adequate for the description of the contribution of
the modulated state since the involved ordered components do not have equal
magnitudes.\ Finally, the above zero-field treatment is obviously not
appropriate for the description of the field-induced metamagnetic phases.

\section{Experimental}

Single crystals of $R$Ni$_{2}$B$_{2}$C ($R$=Er-Gd) were grown by floating zone
method.\cite{Takeya-FZ-method} Physical characterization were carried out
utilizing a dc magnetometer [$H$%
$<$%
50 kOe, 2 K
$<$%
$T$%
$<$%
20 K], a mutual-induction ac-susceptometer [$\nu$=250-500 Hz, $H_{pp}\approx$1
Oe, 30 mK
$<$%
$T$%
$<$%
20 K], and a dc resistivity set-up [$H$%
$<$%
50 kOe, 2 K
$<$%
$T$%
$<$%
20 K]. Structural, magnetic and transport characterizations are in agreement
with published results.\ 

The temperature-dependent specific heat was measured on a semi-adiabatic
calorimeter [80 mK
$<$%
$T$%
$<$%
25 K, precision better than 4\%]. The total specific heat \ curves measured
above 2 K are in agreement with the reported
data.\cite{Canfield-HoNi2B2C-M-Cp,Cho-ErNi2B2C-ansitropy,Tommy-Dy-superconductivity,Tommy-Tb-reorientation}
However, we observed some discrepancy between the absolute values of
$C_{M}(T)$ of single crystal and polycrystalline samples: though both specific
heats were found to be given by approximately the same functional form, the
absolute values of the fit parameters ($\theta$ and $\Delta$) differ by as
much as 40\%.

For each compound,\ the total specific heat, $C_{tot}$, was analyzed as a sum
of an electronic $C_{e}$ ($C_{S}$ when superconductivity is to be emphasized),
a Debye $C_{D}$ $(=\beta T^{3})$, a nuclear $C_{N}$, and a magnetic
contribution $C_{M}$ from the only magnetically active $R$-sublattice. At
temperatures of interest, $C_{e}$ and $C_{D}$ were estimated based on our
specific heat characterization \cite{Takeya-YNi2B2C} of single crystal
YNi$_{2}$B$_{2}$C ($\gamma=17.5$ mJ/moleK$^{2}$ and $\beta=$ 0.12
mJ/moleK$^{4}$) which had been synthesized by the very same procedures as the
one used for the other single crystals. Thus for each compound, $\gamma$ was
taken to be the same value as that of YNi$_{2}$B$_{2}$C while $\beta$ was
evaluated by normalizing the $\beta$ of YNi$_{2}$B$_{2}$C using the
approximation $\beta\propto m^{\frac{3}{2}}$(see Tab.II).

Within the superconducting region, $C_{S}$ was evaluated as\ 3$\gamma
T^{3}/T_{c}^{2}$. At any rate, for all the studied compounds, $C_{M}(T)$ is
much larger than the sum of $C_{e}$ and $C_{D}$. Consequently, even if $C_{e}$
and $C_{D}$ are taken as the bare values of YNi$_{2}$B$_{2}$C, $C_{M}(T)$
would not be noticeably modified, ensuring that our conclusions would not be influenced.

$C_{N}(T)$ of $R$Ni$_{2}$B$_{2}$C, when available, is of \ dominant importance
only at very low temperatures and was evaluated by least-square fit using the
appropriate hyperfine Hamiltonian; the obtained parameters (shown in
Tab.II)\ compare favorably with those of the corresponding $R$-metal
\cite{Anderson-Nclr-Nd-Sm-DY-Schottky} and $R$Co$_{2}$B$_{2}$C
isomorphs.\cite{Massalami-RCo2B2C}

\section{Results and Analysis}

From the general feature of $C_{M}(T)$ curves (Figs.1-3,5-6), one
distinguishes four temperature regions: (i) a paramagnetic region, $T>T_{N}$,
wherein $C_{M}(T)$ is predominantly due to a change in the population of the
CEF levels, (ii) a critical region, $T\approx T_{N}$, wherein $C_{M}(T)$ is
related to the critical phenomena, (iii) an intermediate region wherein
$C_{M}(T)$ reflects the magnetic character of the spiral/modulated states, and
(iv) the low-temperature AF/squared-up states, of prime interest to this work,
wherein the measured $C_{M}(T)$ is to be confronted with Eqs.\ref{Cm-bessel}%
,\ref{Cm-2d} and therefrom $\ \theta$ and $\Delta$ are to be extracted.

Before we discuss the features of $C_{M}(T)$ for each compound, a word of
caution is in order: just as in the case of $R$-metals,\cite{Coqblin-book} the
propagation of errors due to successive subtraction of $C_{e}(T)$, $C_{D}(T)$
and $C_{N}(T)$ would eventually influence the absolute value of $C_{M}(T)$.

\subsection{GdNi$_{2}$B$_{2}$C}

Below $T_{N}$, the zero-field magnetic structure \cite{Detlefs-GdNi2B2C-XRES}%
\ is a transverse sine-modulated type with $\overrightarrow{k_{a}}$ that
changes from 0.551$\overset{\ast}{a}$ at $T_{N}$ to 0.550$\overset{\ast}{a}$
at $T_{R}$, where a spin reorientation process sets-in. Below $T_{R}$,
$\overrightarrow{k_{a}}$ reverts course and increases monotonically till
reaching 0.553$\overset{\ast}{a}$ at 3.5 K.

$C_{M}(T)$ of GdNi$_{2}$B$_{2}$C (Fig.1) reveals the onset of the magnetic
order at $T_{N}=$19.5 K and the spin reorientation process at $T_{R}=$13.5 K,
in agreement with earlier studies.
\cite{Canfield-GdNi2B2C,Detlefs-GdNi2B2C-XRES,Tomala-GdNi2B2C-MES,Godart-GdNi2B2C-composition,Massalami-GdNi2B2C-GdNiBC}
The thermal evolution of $C_{M}(T)$ within the amplitude-modulated state is
distinctly different from that within the equal-amplitude, low-temperature
state (see below). As mentioned above, within the modulated region the
linearized spin-wave analysis is not applicable and one should resort to the
findings of Schmitt and co-workers \cite{Bouvier-Gd-AM-state-Cm-1}. In
particular, these authors demonstrated that $C_{M}(T_{N})$ of such a state
suffers a strong reduction (almost 1/3) in comparison with the value (20.15
J/moleK) expected for an equal-amplitude AF state\ (our results are in
excellent agreement with this statement).%

\begin{figure}
[ptbh]
\begin{center}
\includegraphics[
height=8.8546cm,
width=12.6943cm
]%
{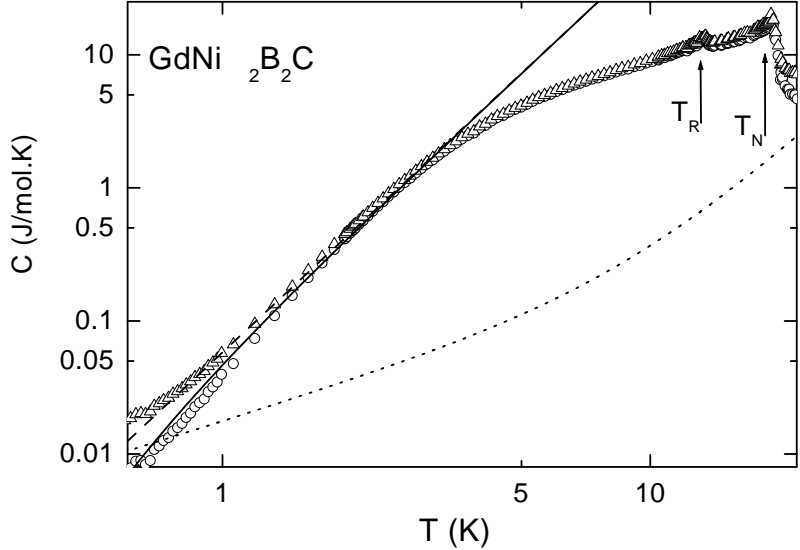}%
\caption{log-log plot of $C_{tot}(T)$ (triangles), $C_{e}(T)+C_{D}(T)$
(dotted), and $C_{M}(T)$ (circles) curves of single crystal GdNi$_{2}$B$_{2}%
$C. The continuous line represents Eq.\ref{Cm-bessel} with $\ \theta=$%
12.5$\pm$0.2 K and $\Delta=$ 1.9$\pm$0.3 K. For $\Delta<T<4$ K, $C_{M}(T)$
follows 0.058$T^{3}$ J/moleK (dashed line) which is the high-T limit of
Eq.\ref{Cm-bessel} (see text).}%
\end{center}
\end{figure}

On the other hand, Fig.1 shows that for temperatures below 3.5 K, $C_{M}(T)$
follows faithfully Eq.\ref{Cm-bessel} with $\theta$=12.5$\pm$0.2K and $\Delta
$=1.9$\pm$0.3 K. \ The numerical value of $\theta$ (for $\Delta$ see Sec.V) is
physically acceptable as can be seen from the following arguments. First, the
substitution of this $\theta$ into Eq.\ref{Theta} yields $J_{eff}$= 0.58$\pm
$0.2 K, which is in close agreement with the value reported for HoNi$_{2}%
$B$_{2}$C (Ref.\cite{Amici-thesis},\cite{Cho-Crystal-field-2}) and TmNi$_{2}%
$B$_{2}$C (Ref.\cite{Movshovich-TmNi2B2C}). Secondly, the substitution of
$\theta$ into Eq.\ref{Cm-AF-highT} predicts correctly the high temperature
limit, namely $C_{M}(T>\Delta)$ =0.058$T^{3}$ J/moleK (see Fig.1). Thirdly,
the substitution of $\theta$ into the Molecular Field
relation:\cite{Kubo-SpinWave}
\begin{equation}
T_{N}=\frac{1}{3}\theta(S+1)\label{Tn-cal}%
\end{equation}
gives $T_{N}$ = 18.8$\pm$0.3 K which is in reasonable agreement with the
experimentally determined value of $T_{N}$.

\subsection{TbNi$_{2}$B$_{2}$C}

A longitudinal SDW, accompanied by an orthorhombic distortion, sets-in at
$T_{N}$%
.\cite{Dervenagas-Tb-structure-anistropy-WeakFM,Song-Tb-dichroism-HRMXRD} The
magnitude of the modulation vector decreases from 0.551$\overset{\ast}{a}$
near $T_{N}$ to 0.545$\overset{\ast}{a}$ at 2.3
K.\cite{Dervenagas-Tb-structure-anistropy-WeakFM} A weak ferromagnetic
component develops below $T_{WF}\approx$ 8 K and at lower temperature a
squaring-up of the modulated state was
observed.\cite{Dervenagas-Tb-structure-anistropy-WeakFM}

$C_{M}(T)$ of single crystal TbNi$_{2}$B$_{2}$C (Fig.2) shows the magnetic
ordering at $T_{N}=14.5$ K and the WF-associated anomaly that peaks around 5.5
K. These features are in agreement with those reported by\ Tommy et
al.\cite{Tommy-Tb-reorientation} \ Similar to the case of GdNi$_{2}$B$_{2}$C,
no attempt was made to analyze $C_{M}(T)$ within the amplitude modulated state
spanning the range $5$ K$<T<T_{N}$. Below 5 K, where the
orthorhombic-distorted squared-up state is expected, $C_{M}(T)$ follows
convincingly the prediction of Eq.\ref{Cm-bessel} with $\theta$=$21.5\pm$0.2 K
and $\Delta$=7.0$\pm$0.5 K.

\bigskip%
\begin{figure}
[ptbh]
\begin{center}
\includegraphics[
height=8.8546cm,
width=12.6943cm
]%
{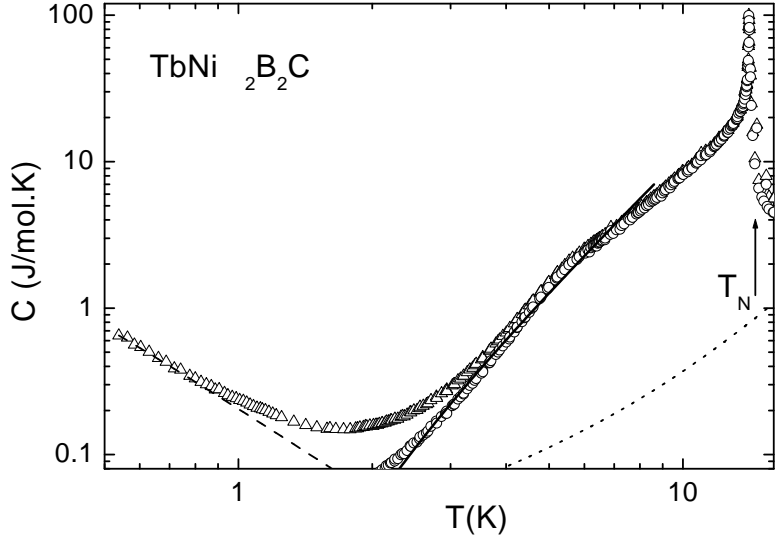}%
\caption{log-log plot of the $C_{tot}(T)$ (triangle), $C_{e}(T)+C_{D}(T)$
(dotted), \ $C_{N}(T)$ (dashed) and \ $C_{M}(T)$ (circle) of single crystal
TbNi$_{2}$B$_{2}$C. The continuous line represents Eq.\ref{Cm-bessel} with
$\Delta=7.0\pm$0.5 K and $\theta=$ 21.5$\pm$0.2 K.}%
\end{center}
\end{figure}

\subsection{DyNi$_{2}$B$_{2}$C}

This compound\ develops a commensurate AF structure below $T_{N}$ with moments
arranged in an identical manner as that of HoNi$_{2}$B$_{2}$C.
\cite{Lynn-RNi2B2C-mag-xrd,Dervenagas-Dy-structure} Moreover,
superconductivity coexists with this AF order below $T_{c}\simeq$6 K. In
contrast to other Ni-based AF superconducting borocarbides, DyNi$_{2}$B$_{2}$C
presents the following distinct features: (i) no zero-field incommensurate or
modulated state was
reported.\cite{Lynn-RNi2B2C-mag-xrd,Dervenagas-Dy-structure} However for $T<$
2 K, anomalously large hysteresis and pronounced reentrant effects were
observed for the field range 1 kOe$\leq H\leq$5.3 kOe,
\cite{Peng-Dy-hystresis-2K} (ii) the superconductivity emerges within a well
developed AF order ($T_{c}<T_{N}$) and that $T_{c}$ is extremely sensitive to
nonmagnetic doping.\cite{Cho-RNi2B2C-deGeness}

$C_{tot}(T)$ (see Fig.3) reveals the onset of the AF order at $T_{N}%
=9.5\pm0.2$ K. Within the accuracy of our measurement, the superconducting
jump at $T_{c}$ $\simeq$6 K is too small to be resolved. On carrying out the
analysis of $C_{tot}$ into its components ($C_{e}$, $C_{D},C_{N}$, and $C_{M}%
$), we observed an anomalous contribution peaking at $1.2$ K and having
features reminiscent of a Schottky-like contribution. Accordingly, it was
approximated by the standard two-level relation:
\begin{equation}
C_{sch}(T)=R(\frac{\delta}{T})^{2}\exp(\frac{\delta}{T})/\left[  1+\exp
(\frac{\delta}{T})\right]  ^{2} \label{Csch}%
\end{equation}
where $\delta$ is the energy separation. It was found out (see the inset of
Fig.3) that $\delta$ $=2.9$ K and that only 0.062 molar fraction was involved.
Moreover, the fit was satisfactory for the high temperature tail but not so
good at the lower temperature part, suggesting that a multi-level Schottky
contribution might be more appropriate. However, for the present discussion,
the above two-level approximation is sufficient. It is highly possible that
such a contribution is due to 6\% defect/impurity which is beyond the limit of
detection of our X-ray structural characterization. Coincidently, anomalous
hystereses effects were observed in the magnetostriction curves that were
measured within the same temperature range.\cite{Peng-Dy-hystresis-2K} At any
case, for $T>\delta$, both $C_{sch}(T)$ and $C_{N}(T)$ are smaller than
$C_{M}(T)$ (see inset of Fig.3). Nevertheless, we considered $C_{M}%
(T)=C_{tot}(T)-C_{S}(T)-C_{D}(T)-C_{N}(T)-C_{sch}(T)$ (see Fig.3).

The thermal evolution of $C_{M}(T)$ is shown in a log-log plot in Fig.3 and as
ln($C_{M}(T)$) $versus$ $1/T$ plot in Fig.4. In both figures, the comparison
with Eq.\ref{Cm-bessel} (solid line) is also presented. Evidently over a wide
range of temperatures, $C_{M}(T\leq T_{c})$ follows excellently
Eq.\ref{Cm-bessel} with $\theta$=19.3$\pm$0.2K and $\Delta$=8.3$\pm$0.3 K.

\bigskip%
\begin{figure}
[ptbh]
\begin{center}
\includegraphics[
height=8.8546cm,
width=12.6943cm
]%
{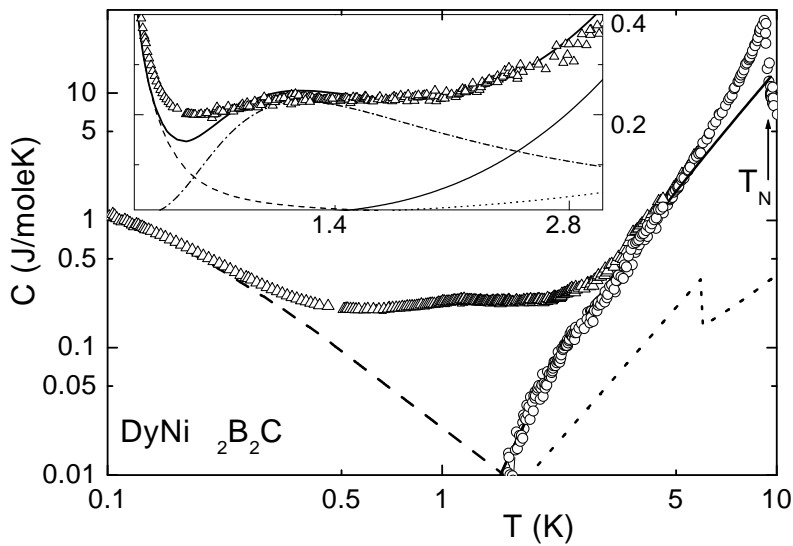}%
\caption{A log-log plot of $C_{tot}(T)$ (triangle), $C_{e}(T)+C_{D}(T)$
(dotted), $C_{N}(T)$ (dashed), and $C_{M}(T)$ (circles) of single crystal
DyNi$_{2}$B$_{2}$C. The solid line represents Eq.\ref{Cm-bessel} (see text).
The inset shows the individual contribution of $C_{tot}(T)$ (symbol),
$C_{N}(T)$ (dashed), the magnetic fit (thin solid line), $C_{sch}(T)$
(dash-dot, see Eq.\ref{Csch}), and the thick solid line is the sum of all
contribution.}%
\end{center}
\end{figure}

%

\begin{figure}
[ptbh]
\begin{center}
\includegraphics[
height=8.8546cm,
width=12.6943cm
]%
{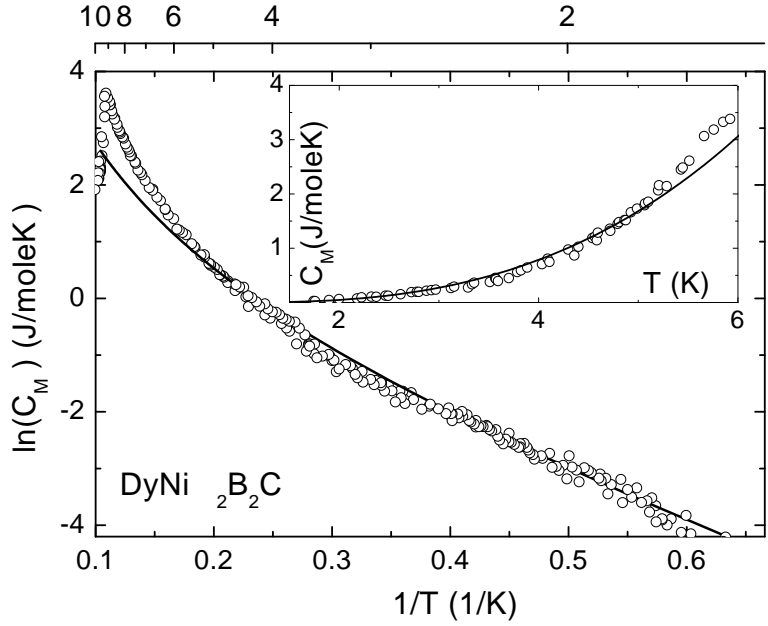}%
\caption{ln($C_{M}(T)$) $versus$ $1/T$ curve \ of single-crystal DyNi$_{2}%
$B$_{2}$C. The data (circles) are compared to Eq.\ref{Cm-bessel} (solid line)
giving $\Delta=$8.3$\pm$0.3 K and $\theta=$ 19.3 $\pm$0.2 K. The inset shows
on a linear scale, $C_{M}(T)$ (circle) together with the comparison to
Eq.\ref{Cm-bessel} (see text).}%
\end{center}
\end{figure}

\subsection{HoNi$_{2}$B$_{2}$C
\ \ \ \ \ \ \ \ \ \ \ \ \ \ \ \ \ \ \ \ \ \ \ \ \ \ \ \ \ \ \ \ \ \ \ \ \ \ \ \ \ \ \ \ \ \ \ \ \ \ \ \ \ \ \ \ \ \ \ \ \ \ \ \ \ \ \ \ \ \ \ \ \ \ \ \ \ \ \ \ \ \ \ \ \ \ \ \ \ \ \ \ \ \ \ \ \ \ \ \ \ \ \ \ \ \ \ \ \ \ \ \ \ \ \ \ \ \ \ \ \ \ \ \ \ \ \ \ \ \ \ \ \ \ \ \ \ \ \ \ \ \ \ \ \ \ \ \ \ \ \ \ \ \ \ \ \ \ \ \ \ \ \ \ \ \ \ \ \ \ \ \ \ \ \ \ \ \ \ \ \ \ \ \ \ \ \ \ \ \ \ \ \ \ \ \ \ \ \ \ \ \ \ \ \ \ \ \ \ \ \ \ \ \ \ \ \ \ \ \ \ \ \ \ \ \ \ \ \ \ \ \ \ \ \ \ \ \ \ \ \ \ \ \ \ \ \ \ \ \ \ \ \ \ \ \ \ \ \ \ \ \ \ \ \ \ \ \ \ \ \ \ \ \ \ \ \ \ \ \ \ \ \ \ \ \ \ \ \ \ \ \ \ \ \ \ \ \ \ \ \ \ \ \ \ \ \ \ \ \ \ \ \ \ \ \ \ \ \ \ \ \ \ \ \ \ \ \ \ \ \ \ \ \ \ \ \ \ \ \ \ \ \ \ \ \ \ \ \ \ \ \ \ \ \ \ \ \ \ \ \ \ \ \ \ \ \ \ \ \ \ \ \ \ \ \ \ \ \ \ \ \ \ \ \ \ \ \ \ \ \ \ \ \ \ \ \ \ \ \ \ \ \ \ \ \ \ \ \ \ \ \ \ \ \ \ \ \ \ \ \ \ \ \ \ \ \ \ \ \ \ \ \ \ \ \ \ \ \ \ \ \ \ \ \ \ \ \ \ \ \ \ \ \ \ \ \ \ \ \ \ \ \ \ \ \ \ \ \ \ \ \ \ \ \ \ \ \ \ \ \ \ \ \ \ \ \ \ \ \ \ \ \ \ \ \ \ \ \ \ \ \ \ \ \ \ \ \ \ \ \ \ \ \ \ \ \ \ \ \ \ \ \ \ \ \ \ \ \ \ \ \ \ \ \ \ \ \ \ \ \ \ \ \ \ \ \ \ \ \ \ \ \ \ \ \ \ \ \ \ \ \ \ \ \ \ \ \ \ \ \ \ \ \ \ \ \ \ \ \ \ \ \ \ \ \ \ \ \ \ \ \ \ \ \ \ \ \ \ \ \ \ \ \ \ \ \ \ \ \ \ \ \ \ \ \ \ \ \ \ \ \ \ \ \ \ \ \ \ \ \ \ \ \ \ \ \ \ \ \ \ \ \ \ \ \ \ \ \ \ \ \ \ \ \ \ \ \ \ \ \ \ \ \ \ \ \ \ \ \ \ \ \ \ \ \ \ \ \ \ \ \ \ \ \ \ \ \ \ \ \ \ \ \ \ \ \ \ \ \ \ \ \ \ \ \ \ \ \ \ \ \ \ \ \ \ \ \ \ \ \ \ \ \ \ \ \ \ \ \ \ \ \ \ \ \ \ \ \ \ \ \ \ \ \ \ \ \ \ \ \ \ \ \ \ \ \ \ \ \ \ \ \ \ \ \ \ \ \ \ \ \ \ \ \ \ \ \ \ \ \ \ \ \ \ \ \ \ \ \ \ \ \ \ \ \ \ \ \ \ \ \ \ \ \ \ \ \ \ \ \ \ \ \ \ \ \ \ \ \ \ \ \ \ \ \ \ \ \ \ \ \ \ \ \ \ \ \ \ \ \ \ \ \ \ \ \ \ \ \ \ \ \ \ \ \ \ \ \ \ \ \ \ \ \ \ \ \ \ \ \ \ \ \ \ \ \ \ \ \ \ \ \ \ \ \ \ \ \ \ \ \ \ \ \ \ \ \ \ }%

$C_{tot}(T)$ of single crystal HoNi$_{2}$B$_{2}$C (shown in Fig.5) reveals a
cascade of three transitions that are usually attributed to magnetic
transformations.\cite{Canfield-HoNi2B2C-M-Cp,Elhagary-HoNi2B2C-Cp} The
signature of the onset of superconductivity is too weak to be observable in
our present measurements. On the other hand, for $T<$5 K, $C_{M}(T)$ follows
the description of Eq.\ref{Cm-bessel} with $\theta$=9.7$\pm$0.2K and $\Delta
$=8.3$\pm$0.3 K.%

\begin{figure}
[ptbh]
\begin{center}
\includegraphics[
height=8.8546cm,
width=12.6943cm
]%
{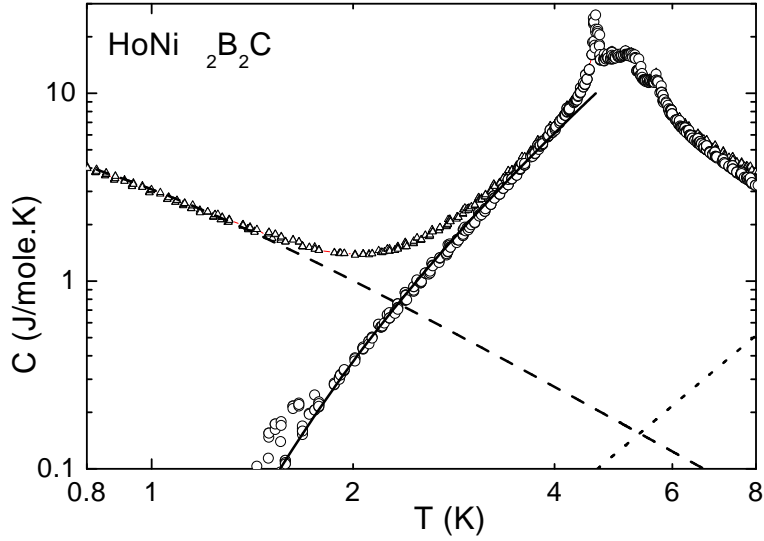}%
\caption{A log-log plot of $C(T)$ $versus$ $T$ of single crystal HoNi$_{2}%
$B$_{2}$C showing $C_{tot}(T)$ (triangle), $C_{e}(T)+C_{D}(T)$ (dotted),
\ $C_{N}(T)$ (dashed), and \ $C_{M}(T)$ (circle) contributions. The magnetic
contribution (circles) are compared to Eq.\ref{Cm-bessel} (solid line) giving
$\Delta=$8.3$\pm$0.3 K and $\theta=$ 9.7$\pm$0.2 K(see text)}%
\end{center}
\end{figure}

\subsection{ErNi$_{2}$B$_{2}$C}

Two intriguing features of the $H-T$ phase diagram of ErNi$_{2}$B$_{2}$C are
$\ $%
\cite{Lynn-RNi2B2C-mag-xrd,Canfield-Er-HT-diagram,Choi-ErNi2B2C-neutron,Kawano-ErNi2B2C-neutron}%
: the onset of the incommensurate transversely polarized SDW state with
\textit{k}$_{a}$=0.553$a^{\ast}$ at $T_{N}=$5.94 K, and the onset of weak
ferromagnetism (WF) at $T_{WF}=$2.2 K. These two events\ (none is able to
quench superconductivity, $T_{c}=10.5$ K) are well evident in $C_{M}(T)$ of
Fig.6. $T_{WF}$, in particular, is evident as a change of slope that separates
two distinct thermal evolutions:\cite{Massalami-Cm-ErNi2B2C} $C_{M}(T)$ within
the amplitude-modulated state $T_{WF}<T<T_{N}$ and that within the squared-up
state at $T<T_{WF}$. In the latter region, $C_{M}(T<T_{WF})$ is well described
by Eq.\ref{Cm-bessel} with $\ \theta=$7.4$\pm$0.2 K and $\Delta=$5.4$\pm$0.3
K. %

\begin{figure}
[ptbh]
\begin{center}
\includegraphics[
height=8.8546cm,
width=12.6943cm
]%
{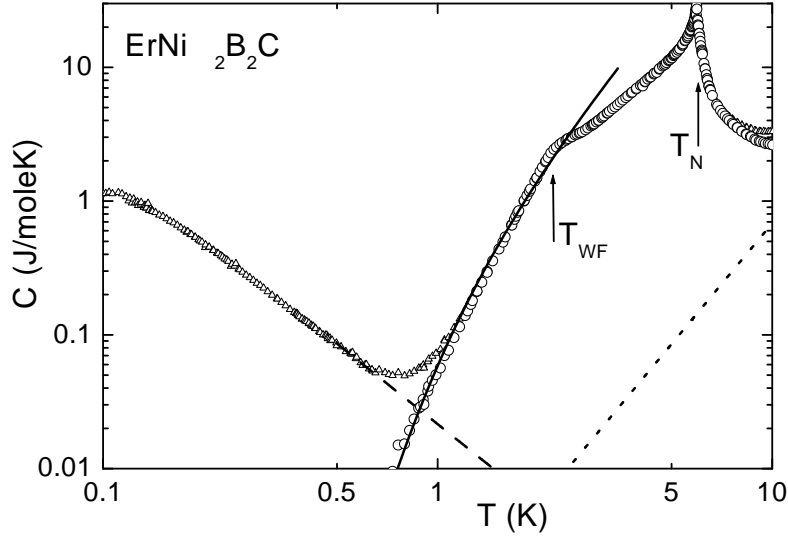}%
\caption{A log-log plot of $C_{tot}(T)$ (triangle), $C_{e}(T)+C_{D}(T)$
(dotted), $C_{N}(T)$ (dashed), and $C_{M}(T)$ (circle) of single crystal
ErNi$_{2}$B$_{2}$C. The solid line is a comparison to Eq. \ref{Cm-bessel} with
$\ \theta=$7.4$\pm$0.2 K and $\Delta=$ 5.4$\pm$0.3 K (see text).}%
\end{center}
\end{figure}

\subsection{TmNi$_{2}$B$_{2}$C}

Superconductivity sets-in at 11 K and, below $T_{N}=1.52\pm0.05$ K, coexists
with a transversely polarized SDW state wherein the spins are pointing along
the c-axis and the modulation vector is
(0.093,0.093,0).\cite{Lynn-RNi2B2C-mag-xrd,Movshovich-TmNi2B2C} At lower
temperatures, the incommensurate SDW state squares
up.\cite{Lynn-RNi2B2C-mag-xrd} The magnetic specific heat of single crystal
TmNi$_{2}$B$_{2}$C was measured by Movshvich et al \cite{Movshovich-TmNi2B2C}
and was shown to follow Eq.\ref{Cm-2d} with $J_{\bot}\approx$0.8 K, $J_{\Vert
}\approx$0.2 K.

\section{Discussion and Conclusions}

On a linearized spin-wave (non interacting magnon gas) approach, one is always
limited to the low-temperature region of the ordered state which in the case
of the borocarbides amounts to being restricted to below the liquid Helium
temperatures. Within that temperature range, the model describes very
successfully the magnetic contribution of the studied compounds indicating
that $C_{M}$ can be safely associated with the gapped collective excitations
that propagate within the orthorhombic-distorted 3d AF (-type) structure.

The successful applicability of the model to the commensurate collinear AF
ground structures of $R=$Ho, Dy is understandable. To justify its
applicability to the cases of $R=$Tm, Er, Tb, Gd, it is sufficient to show
that their ground structures are well squared-up and that all the moments do
have equal amplitudes and orient \ (or bunch) along a specific direction. The
state of ErNi$_{2}$B$_{2}$C below $T_{WF}$ and that of GdNi$_{2}$B$_{2}$C
below 3.5 K provide the best illustrations of the fulfillment of this
requirement. The colinear, equal-amplitude, and squared-up character of \ the
state of ErNi$_{2}$B$_{2}$C below $T_{WF}$ was elegantly revealed in the
neutron diffractions studies of Choi et al\cite{Choi-ErNi2B2C-neutron} and
Kwano-Furukawa et al \cite{Kawano-ErNi2B2C-neutron}. It is remarked that the
presence of \ weak ferromagnetism (reflected as kink that separate oppositely
oriented domains) would hardly modify this picture since the excitation energy
of the kink is much higher than that of the magnon. Let us now discuss the
character of the state of GdNi$_{2}$B$_{2}$C below 3.5 K. A recent
magnetoelastic study \cite{Massalami-GdNi2B2C-singlecrystal-1} on single
crystal GdNi$_{2}$B$_{2}$C demonstrated the presence of substantial
magnetoelastic and anisotropic exchange interactions, in particular below
$T<T_{R}$. The magnitude of the $\epsilon^{\gamma}$ strain mode (which leads
to symmetry lowering from four-fold tetragonal to two-fold orthorhombic) is
very large and increases with decreasing temperature leading progressively to
an orthorhombic-distorted magnetic state wherein the moments, due to entropy
arguments, approach equal amplitudes. Based on the suggestion of Detlefs et al
\cite{Detlefs-GdNi2B2C-XRES}, the magnetic structure below $T_{R}$ \ is either
a \ transverse modulated state with moment orientation away from the b-axis in
the bc-plane or a modified spiral-like structure which, due to the low
symmetry (2mm) of the [100] direction, is likely to suffer fanning or bunching
(becoming more stronger as the temperature is decreased). Tomala et al
\cite{Tomala-GdNi2B2C-MES}, on exploring these two structural possibilities,
argued that the $^{155}$Gd Mossbauer spectra at 4.2K
$<$%
$T<T_{R}$ were better fitted with a bunched spiral-like state. In this
Mossbauer analysis, the c-component was found to be almost equal ( $\sim$75\%
at 4.2 K) to the b-component, a result that does not reflect the 2mm-symmetry
of the [100] direction. Considering these and the above observations together
with entropy arguments, one concludes that the low-temperature structure is
either a \emph{squared-up}, \emph{equal-amplitude}, and collinear state or -
as supported by Mossbauer analysis- a strongly bunched, \emph{squared-up}, and
\emph{equal-amplitude} state. As far as the magnon specific heat is concerned,
the interactions in both structures can be represented by the Hamiltonian of
Eq.\ref{H} (allowing for a unimportant difference in the direction of the axis
of quantization).

Figs.1-6 and Tab.I demonstrated convincingly that, based on only $\Delta$ and
$\theta,$ the diverse functional form of the measured $C_{M}(T)$ can be
systematized: when both $\Delta$ and $\theta$ are large, $C_{M}(T<\Delta)$
reflects a\ magnon contribution from an anisotropic magnetic structure as in
$R=$Er, Ho, Dy, Tb. When $\Delta$ is relatively small but $\theta$ is large,
$C_{M}(T>\Delta)$ reflects a magnetic contribution from a quasi-isotropic
magnetic structure as in GdNi$_{2}$B$_{2}$C. For a weak $\Delta$ and
$J_{\Vert}$, $C_{M}(T)$ reflects a contribution from a quasi-2$d$ structure as
in TmNi$_{2}$B$_{2}$C.

The evolution of $\Delta$ and $\theta$ across the studied compounds is
reasonable. $\theta$, on the one hand, reflects predominately the evolution of
the de Gennes factor (see Eqs.5,7 and Tab.I) as can be appreciated on
observing that $\theta$ scales very well with the de Gennes factors for
$R$=Tm, Er, Ho, Dy, Tb. That the experimentally determined $\theta$ of
GdNi$_{2}$B$_{2}$C is a factor of three lower than the one expected from de
Gennes scaling may be attributed to the additional dependence of $\theta$ on
$H_{a}$ which for GdNi$_{2}$B$_{2}$C is the lowest.

$\Delta$, on the other hand, reflects the combined influence of the
anisotropic forces and interlayer coupling. This is expressed by Eq.\ref{gap}
which for the case of, say, ErNi$_{2}$B$_{2}$C (considering $H_{A}$ $\approx
$15 kOe and $\left\vert \ J_{\Vert}\right\vert \approx$0.1 K) gives a value of
4 K which is close to the experimental value. The observation that $\Delta$ is
nonzero for each of the studied compounds is in agreement with the reported
anisotropic features of the magnetic and transport
properties.\cite{Cho-ErNi2B2C-ansitropy} The strong anisotropy of \ each of
$R=$Er, Ho, Dy, Tb is in accord with what is expected from their CEF
properties. In contrast, the weak anisotropy observed in GdNi$_{2}$B$_{2}$C is
most probably due to a combination of anisotropic exchange, dipolar, and
two-ion magnetoelastic couplings.

It is interesting to discuss one particular aspect of \ the interaction
between magnons and superconductivity in, say, $R$= Ho ($T_{N}<T_{c}$) and Dy
($T_{c}<T_{N}$): the magnon-mediated pairbreaking effect that manifests itself
in the thermal evolution of $H_{c2}(T<T_{N},T_{c})$
[Ref.\cite{Mag-Sup-Fischer-1990}]. Noteworthy, the thermal evolution of
$H_{c2}(T<T_{N},T_{c})$ of \ both HoNi$_{2}$B$_{2}$C and DyNi$_{2}$B$_{2}$C
[Ref.\cite{Canfield-RNi2B2C-Hc2}] are very similar which, considering the
above-mentioned similarity in their magnetic properties, suggests that the
involved pairbreaking effects (in particular the magnon-mediated one) are
similar. This, in turn, suggests that the magnon characteristic (say
low-energy magnon spectra) in both compounds must be similar. This is indeed
the case: the analysis of Sec.IV.C-D showed that the energy cost for magnon
excitation in both compounds is practically equal ($\Delta\simeq$ 8 K see
Tab.I). Therefore doping of Ho into DyNi$_{2}$B$_{2}$C (up to 80\% but still
$T_{c}<T_{N}$) would not lead to a softening of $\Delta$. Then, for this
concentration range, there should be no variation in $H_{c2}$ and $T_{c}$ even
though the deGennes factor does vary (remember that within the
antiferromagnetic state, the Abrikosov-Gorkov pairbreaking mechanism is not
valid). This provides an additional experimental confirmation of the
hypothesis of Cho et al \cite{Cho-RNi2B2C-deGeness} that the magnon spectrum
of (Dy$_{1-x}$Ho$_{x}$)Ni$_{2}$B$_{2}$C is hardly modified for $x<$0.8. In
contrast, for (Dy$_{1-x}$Ho$_{x}$)Ni$_{2}$B$_{2}$C ($x>$0.8),\ the onset of
superconductivity occurs within the paramagnetic state and consequently the Dy
dopant depresses $T_{c}$ linearly as expected from the Abrikosov-Gorkov
theory.\medskip

In summary, we were able to reveal the magnon specific heat contribution of
the heavy members of the borocarbides and to identify the expressions that
describe their thermal evolution. These expressions (given in terms of only
two physically accepted parameters) were derived from the spin-wave analysis
of a simple Hamiltonian\ that consists of effective exchange couplings and
anisotropic interactions. We investigated as well an influence of the magnons
on the superconductivity of these AF superconductors.

Improvements and extension of this analysis are underway. These include, on
the experimental side, probing the magnon contribution in single crystals of
$R$Ni$_{2}$B$_{2}$C by other (microscopic and macroscopic) techniques and, on
the theoretical side, a better and more realistic approximation of \ $J(k)$,
CEF effects, and magnetoelastic, and anisotropic exchange forces.

\acknowledgments Partial financial support were provided by Brazilian agencies
CNPq and FAPERJ.

\newpage

\subsection{Tables}

TAB.I \ Some zero-field parameters of \ selected $R$Ni$_{2}$B$_{2}$C
compounds. Superconducting $T_{c}$, magnetic $T_{N}$, \textit{magnetic
structure}, propagation \textit{wave vector}, and \textit{moment direction}
are taken from Ref. \cite{Lynn-RNi2B2C-mag-xrd}. Squaring of the modulated SDW
state is taken to occur at lower temperatures. The gap, $\Delta,$ and
characteristic temperature, $\theta,$ were determined from the indicated
equation and figure. $\theta_{\exp}$ of TmNi$_{2}$B$_{2}$C is calculated by
substituting into Eqs.\ref{C-definition},\ref{Theta} \ the fit values
($J_{\bot}$=.8 K and $\left\vert J_{\Vert}\right\vert $=.2) given by Movshvich
et al.\cite{Movshovich-TmNi2B2C}. $\theta_{deG}$ is the de Gennes scaling of
$\theta$ taking that of $R=$Ho as a reference.%

\begin{tabular}
[c]{|l|l|l|l|l|l|l|l|l|l|l|l|}\hline
$R$ & deG & $T_{c}$ & $T_{N}$ & magnetic & wave & moment & $\Delta$ &
$\theta_{\exp}$ & $\theta_{deG}$ & Eq.\# & Fig.\#\\
&  & K & K & structure & vector & direction & $\pm$0.2 K & $\pm$0.2 K & K &  &
\\\hline
Gd & 15.75 & 0 & 19.5 & SDW & [.55,0,0] & [0,1,0] & 1.9 & 12.5 & 34.0 &
\ref{Cm-bessel} & 1\\\hline
Tb & 10.5 & 0 & 15.4 & SDW & [.555,0,0] & [1,0,0] & 7 & 21.5 & 22.6 &
\ref{Cm-bessel} & 2\\\hline
Dy & 7.08 & 6 & 9.5 & 3d, AF & [0,0,1] & [1,1,0] & 8.3 & 19.3 & 15.3 &
\ref{Cm-bessel} & 3\\\hline
Ho & 4.5 & 8 & 5 & 3d, AF & [0,0,1] & [1,1,0] & 8.3 & 9.7 & 9.7 &
\ref{Cm-bessel} & 5\\\hline
Er & 2.5 & 10.5 & 5.9 & SDW & [.553,0,0] & [0,1,0] & 5.4 & 7.4 & 5.5 &
\ref{Cm-bessel} & 6\\\hline
Tm & 1.16 & 11 & 1.5 & SDW & [.093,.093,0] & [0,0,1] & 0 & 3.8 & 2.5 &
\ref{Cm-2d} & 4 in Ref.\cite{Movshovich-TmNi2B2C}\\\hline
\end{tabular}

\bigskip

\bigskip

TAB.II List of values of $\gamma$, $\beta,$ and $\theta$ of $R$Ni$_{2}$B$_{2}%
$C ($R$=Gd,Tb, Er, Ho, Dy, Tm). \ Also given are the expressions used for
evaluating $C_{s}(T<T_{c}).$ The nuclear hyperfine parameters $\alpha$ and $P$
(see Refs.\cite{Anderson-Nclr-Nd-Sm-DY-Schottky}) are also indicated: The
upper (lower) values of DyNi$_{2}$B$_{2}$C correspond to the isotope $^{161}%
$Dy ($^{163}$Dy). Data of TmNi$_{2}$B$_{2}$C were taken from
Ref.\cite{Movshovich-TmNi2B2C}.%

\begin{tabular}
[c]{|l|l|l|l|l|l|l|}\hline
$R$ & $\gamma$ & $C_{s}$ & $\theta$ ($\beta)$ & \multicolumn{2}{|l|}{$C_{N}$}
& Fig.\#\\
& mJ/moleK$^{2}$ & J/moleK & K (mJ/moleK$^{4}$) & $\alpha$ (K) & $P$(K) &
\\\hline
Gd & 17.5 & $\gamma T$ & 392(0.1935) & 0 & 0 & 1\\\hline
Tb & 17.5 & $\gamma T$ & 391(0.196) & 0.14(2) & 0.02(1) & 2\\\hline
Dy & 17.5 & 3$\gamma T^{3}/T_{c}^{2}$ & 388 (0.200) & $%
\begin{tabular}
[c]{l}%
$-.0396$\\
.0554
\end{tabular}
\ $ & $%
\begin{tabular}
[c]{l}%
$.009$\\
.01
\end{tabular}
\ $ & 3\\\hline
Ho & 17.5 & 3$\gamma T^{3}/T_{c}^{2}$ & 386 (0.203) & $0.29$ & $.009$ &
5\\\hline
Er & 17.5 & 3$\gamma T^{3}/T_{c}^{2}$ & 384 (0.206) & $0.045$ & $-.0001$ &
6\\\hline
Tm & $\sim$18 & $\gamma T$ & $\sim$355($\sim$0.26) & $0.0202$ & $0$ & 4 in
Ref.\cite{Movshovich-TmNi2B2C}\\\hline
\end{tabular}

\end{document}